\documentclass[12pt,oneside, a4paper]{article}

\pdfoutput=1  
 
\ifx\pdfoutput\undefined
\usepackage[dvips,bookmarks=false]{hyperref}	
\else
\usepackage{hyperref}	
\fi
\hypersetup{colorlinks,bookmarksopen,bookmarksnumbered,citecolor=blue,
linkcolor=black,pdfstartview=FitH,urlcolor=blue}

\usepackage{graphicx}
\usepackage{subfigure}
\usepackage{amssymb}
\usepackage{cite}
\usepackage{bm}
\usepackage{amsmath,amsthm}
\usepackage{cleveref}

\numberwithin{equation}{section}

\newcommand{\gag}{g_{a\gamma}}

\begin{document}


\begin{center}

\vspace*{2cm}
        {\Large\bf Comment on ``Axion Electrodynamics in the Presence of Current Sources''}
\vspace{1cm}

\renewcommand{\thefootnote}{\fnsymbol{footnote}}
{\bf Thomas Schwetz}$^a$\footnote[3]{schwetz@kit.edu},
{\bf Elisa Todarello}$^{b,c}$\footnote[4]{elisamaria.todarello@unito.it} 
\vspace{5mm}

{\it%
$^a$ Institut f\"ur Astroteilchenphysik, Karlsruhe Institute of Technology (KIT),\\ 76021 Karlsruhe, Germany \\
$^b$  Dipartimento di Fisica, Università di Torino, Via P. Giuria 1, \\10125 Torino, Italy\\
$^c$ Istituto Nazionale di Fisica Nucleare, Sezione di Torino, Via P. Giuria 1,\\ 10125 Torino, Italy}


\abstract{In this note we refute the critique raised in a preprint by J.~Berger and A.~Bhoonah~\cite{Berger:2023muj} on the method used in our earlier paper, Beutter et al.~\cite{Beutter:2018xfx}, to calculate the electromagnetic fields induced by an axion background in the presence of a magnetic field.}

\end{center}

\renewcommand{\thefootnote}{\arabic{footnote}}
\setcounter{footnote}{0}


\section{Introduction}

In our previous paper ``Axion-electrodynamics: a quantum field calculation'' \cite{Beutter:2018xfx} we have discussed a method to calculate electromagnetic fields induced by an axion background in an external magnetic field, adopting methods based on quantum field theory (QFT). This approach has been criticized in the recent article by J.~Berger and A.~Bhoonah \cite{Berger:2023muj} (BB in the following). There are basically two issues raised by BB \cite{Berger:2023muj}:
\begin{enumerate}
\item In \cite{Beutter:2018xfx} we use the Feynman propagator, as common in standard QFT calculations of amplitudes for cross sections, decay rates, and similar. This has been criticized by BB: for classical sources, the retarded propagator/Greens function should be used; the Feynman propagator can lead to acausal results.

\item BB claim that our ``E-field solution does not satisfy the axion electrodynamics version of Maxwell’s equations, particularly Amp\'ere’s law''.
\end{enumerate}

In this note, we are going to address these two points. We partially agree with the first point. However, we show below that it is irrelevant for the situation of interest, and in particular, it does not change the conclusion that the induced electric field in a small experiment is parametrically suppressed by a factor $(m_aR)^2$, where $m_a$ is the axion mass and $R$ is the size of the experiment. 

We do not agree with the second point. In \cref{sec:maxwell} we show that the solution obtained in \cite{Beutter:2018xfx} for axion-induced electric and magnetic fields in the case of an infinitely long solenoid indeed does satisfy the full set of Maxwell's equations coupled to an axion at leading order in the coupling constant.

\section{Feynman and retarded propagators}

We revisit the calculation for the axion-dark matter-induced electromagnetic (EM) fields of \cite{Beutter:2018xfx}. We consider the case of zero-velocity axions:
\begin{equation}\label{eq:axion}
  a(y) = a_0 e^{-im_a y_0} \,. 
\end{equation}
The induced EM potential according to \cite{Beutter:2018xfx} is then
\begin{equation}
  A_\mu^{\rm ind}(x) = ig_{a\gamma} (-im_aa_0) \int d^4y \, e^{-im_a y_0} F^{0\nu}(y)
  \int \frac{d^4q}{(2\pi)^4} \frac{-ig_{\mu\nu}}{q^2} e^{-iq(x-y)} \,.
\end{equation}
For the external magnetic field, we factorize the time dependence $f(y_0)$ and assume a top-hat in
Fourier space:
\begin{equation}
  \int d^3y F_{0i}(y)e^{-i\vec q\cdot\vec y} = R^3 B_i f(y_0) \Theta(Q-|\vec q\,|) \,,
\end{equation}
where $R=1/Q$ corresponds to the typical size of the experiment and $B_i$ is a constant vector with dimension of a magnetic field, whereas $f(y_0)$ is dimensionless.
We introduce the abbreviation $|\vec q| \equiv q$ (not to be confused with the 4-vector).
With this ansatz for the external field we obtain for the spatial components of the induced vector potential
\begin{align}
  \vec A^{\rm ind}(x) &= \frac{-i}{(2\pi)^3} g_{a\gamma} a_0 m_a R^3 \vec B  \int_0^Q dq \,q^2 \,
  \frac{2  \sin(q |\vec x|)}{q|\vec x|}  \, I \,, \label{eq:Aind}\\
  I &= \int dy_0 \, f(y_0)\,e^{-im_a y_0} \int dq_0  \frac{e^{-iq_0(x_0-y_0)}}
  {(q_0 -  q + i\epsilon)(q_0 + q \pm i\epsilon)}  \,. \label{eq:I}
\end{align}
We introduced the imaginary $i\epsilon$ to shift the poles from the real axis, where the limit $\epsilon\to 0^+$ is understood. If $x_0-y_0$ is positive (negative) we have to close the contour below (above). Hence for the upper sign in the last term, we pick up both residues for $x_0 > y_0$ and get zero for $x_0<y_0$, which corresponds to the retarded propagator, whereas for the lower sign in the last term we obtain the Feynman propagator used in standard QFT amplitude calculations. 

Let us investigate now in some detail the integral in \cref{eq:I}. We use the notation $I^{F}$ ($I^R$) to denote \cref{eq:I} evaluated using the Feynman (retarded) propagator.

\subsection{Stationary source}

We start with the situation considered in \cite{Beutter:2018xfx},
namely a stationary source, $f(y_0)=1$, and the Feynman propagator. We can calculate the integral now in two ways. First, we start with the $y_0$ integration, giving a $\delta$-function, which can be directly evaluated:
\begin{align}
  I^{F} &= 2\pi\int dq_0 \, \delta(m_a-q_0) \frac{e^{-iq_0x_0}}{(q_0-q +i\epsilon)(q_0+q -i\epsilon)} \\
  &=2\pi\, \frac{e^{-im_a x_0}}{m_a^2-q^2} \,.
\label{eq:IF}
\end{align}
We note that the imaginary shift of the poles becomes irrelevant and the result is
independent of which type of propagator we use, $I^{F}=I^{R}$.

Second, in order to check explicitly the effect of the different pole descriptions for Feynman versus retarded propagators, let us exchange the order of integration and use contour integration to perform the $q_0$ integral:
\begin{align}
  I^{F} &= \int dy_0 \,e^{-im_a y_0} \frac{-2\pi i}{2q} \left[
    e^{-iq(x_0-y_0)}\Theta(x_0-y_0) + e^{iq(x_0-y_0)}\Theta(y_0-x_0) \right] \\
  &= -\frac{i\pi}{q} e^{-im_ax_0} \int dt
  \left[ \Theta(t)e^{i(m_a-q)t} + \Theta(t)e^{-i(m_a+q)t} \right]
\end{align}
where in the second line we performed a suitable shift of the integration variable.
Hence we need to perform a Fourier transform of the $\Theta$-function. A convenient expression is
\begin{align}\label{eq:Theta}
  \int dt \, e^{-i\omega t} \Theta(t) = 
    -\frac{i}{\omega - i\epsilon}  \qquad (\epsilon \to 0^+)\,,
\end{align}
based on the integral-representation of the $\Theta$-function,\footnote{
Another common version of the Fourier transform is given 
by $\int dt \, e^{-i\omega t} \Theta(t) = 
    \pi \delta(\omega) -  i \mathcal{P} (1/\omega)$, where $\mathcal{P}$ indicates that when this expression appears under an integral, the Cauchy principal value is understood.
The  Sokhotski–Plemelj theorem~\cite{SP-theorem} guarantees that these two expressions provide the same result, once they are integrated over.}
which leads to
\begin{align}  
  I^{F} &= -\frac{i\pi}{q} e^{-imx_0}
  \left(\frac{i}{m_a-q} -\frac{i}{m_a+q} \right) \,, \label{eq:IF2a}
\end{align}
which is identical to \cref{eq:IF} (as it should be).


Let us now consider the retarded propagator, but still assume the stationary source. As mentioned above, if we first perform the $y_0$ integral, the choice of how to shift poles is irrelevant, and we recover the same result as for Feynman. Let's check the calculation by doing first the $q_0$ integral. We pick up both poles and obtain an overall $\Theta$-function:
\begin{align}
  I^{R} &= - \frac{2\pi i}{2q} \int dy_0\, e^{-im_a y_0}\left[ e^{-iq(x_0-y_0)}-e^{iq(x_0-y_0)} \right]
  \Theta(x_0-y_0) \\
  &= - \frac{\pi i}{q} e^{-im_a x_0} \int dt\, \Theta(t) \left[ e^{-i(q-m)t}-e^{i(q+m)t} \right] \\
  &= - \frac{\pi i}{q} e^{-im_a x_0}  \left[ -\frac{i}{q-m} -\frac{i}{q+m} \right] \\
  &=2\pi\, \frac{e^{-im_a x_0}}{m_a^2-q^2}  \,.
\end{align}

We conclude that for the stationary source, Feynman and retarded propagators give the same result.

\subsection{Non-stationary source}

Let's repeat the calculation for a source switched on at $t=0$ by setting $f(y_0) = \Theta(y_0)$. We calculate the integral $I$ from \cref{eq:I} by performing first the $y_0$ integration:
\begin{align}
  I &= \int dq_0 \frac{e^{-iq_0x_0}}
  {(q_0 -  q + i\epsilon)(q_0 + q \pm i\epsilon)}
  \int dy_0 \, \Theta(y_0)\,e^{-i(m_a-q_0)y_0} \\
  &= i \int dq_0 \frac{e^{-iq_0x_0}}
  {(q_0 - m_a +i\epsilon)(q_0 -  q + i\epsilon)(q_0 + q \pm i\epsilon)} \,,
\end{align}
where again we used \cref{eq:Theta} for the Fourier transform of the $\Theta$-function. The expression in the $q_0$ integral has three poles, leading to three terms from the residues. We write $I = \bar I + \tilde I$, where $\bar I$ denotes the term from the residue at $q_0=m_a$
whereas $\tilde I$ includes the terms from the other two poles at $q_0 = \pm q$. We find
\begin{align}
  \bar I &=2\pi\, \frac{e^{-im_a x_0}}{m_a^2-q^2} \Theta(x_0) \,,
\end{align}
i.e., the pole at $q_0=m_a$ gives the same expression as the stationary source multiplied with $\Theta(x_0)$. Depending on the sign combinations the contributions from the other two poles are of the following form:
\begin{align}
  \tilde I \supset \pm \frac{e^{\pm i q x_0}}{2q(q\pm m_a)}\Theta(\pm x_0) \,.
\end{align}

Let us now discuss these results:
\begin{itemize}
\item For the retarded propagator all $i\epsilon$ terms have a positive sign. Hence, we pick up the residues only when closing the contour below, which we have to do for $x_0>0$. Therefore, in this case, all terms will be proportional to $\Theta(x_0)$, as required by causality: the induced field appears only for $t>0$, when the source is switched on.
\item For the Feynman propagator also a term proportional to $\Theta(-x_0)$ appears, apparently violating causality. The reason is the following: The origin of the Feynman propagator is the time-ordered product to be used in the perturbative treatment of ``short'' interactions, where the interaction happens in a finite volume of space-time and at $t=\pm\infty$ all involved states are asymptotically free. This situation typically does not apply in the presence of a classical source as considered here.  
\item
  Let's consider the $q$ integral of the terms in $\tilde I$, again focusing on the center of the experiment at $\vec x = 0$. We find that\footnote{In the case of a small experiment ($Q\gg m_a$) the integration domain may include a pole at $m_a=q$. It can be shown that the contribution of the pole is finite and at leading order, the result is given by \cref{eq:Itilde}.}
  \begin{align}\label{eq:Itilde}
    \int_0^Q dq\, q^2 \tilde I \propto \int_0^Q dq \frac{q}{q\pm m_a} e^{\pm iqx_0}
    \sim
\left\{
\begin{array}{l@{\quad}l}
    \frac{1}{x_0} e^{\pm iQx_0} & (Q \gg m_a)\\
    \frac{Q}{m_a x_0} e^{\pm iQx_0} ,\, \frac{1}{m_a x_0^2} e^{\pm iQx_0} & (Q \ll m_a)\\  
\end{array} \right.
  \end{align}
In all cases, we find that the contributions from $\tilde I$ vanish in the large time limit as $1/x_0$, which holds also in the general case, without assuming any particular relation between $Q$ and $m$.
Hence, these terms describe the time-dependent effects of switching on the source, but they become irrelevant at late times and the result converges to the solution for the stationary source. This behavior appears both for the retarded and the Feynman propagator, and therefore also the issues with causality violation for the Feynman case disappear for large $|x_0|$.
\end{itemize}

In summary, for the Feynman propagator causality is violated for ``short times'' around switching on the experiment; for times much earlier than the instant of switching on the experiment the induced field vanishes. ``Short times'' can be defined by comparing the contributions from $\bar I$ and $\tilde I$. If we consider the induced electric field, we need to take the time derivative of the potential, which gives another factor of $m_a$ ($Q$) from the $\bar I$ ($\tilde I$) contributions, and we find that for large $x_0$ the $\tilde I$ contribution is suppressed to the one from $\bar I$ by $1/(x_0 Q)$ for $Q\ll m_a$ and by $1/(m_ax_0)$ for $m_a\ll Q$. 

Hence, for times sufficiently long after switching on the experiment, we recover the result for the static source, the induced fields become independent of the used propagator and are in agreement with the results obtained in \cite{Beutter:2018xfx}.

\section{The induced EM fields do satisfy Maxwell's equations}
\label{sec:maxwell}

Let us now address the second point of BB~\cite{Berger:2023muj} and show that the solution obtained by the QFT method in \cite{Beutter:2018xfx} indeed does satisfy the full set of Maxwell's equations coupled to an axion background field. We consider the case of an infinitely long solenoid with radius $R$, implying a constant magnetic field $B_0$ in $z$-direction in the inside region $r<R$. We work in cylindrical coordinates. We obtain the induced field in the case of zero axion velocity, taking the axion field to be $a(t) = a_0\,e^{-imt}$.

For convenience we reproduce here the solution for the induced electric and magnetic fields obtained in \cite{Beutter:2018xfx}, eqs.~(3.4) and (3.5) therein. The non-zero components are
\begin{align}
E^z_\mathrm{ind} &= - \gag a_0 B_0 e^{-im_a t} 
\begin{cases}
\displaystyle 1 - \frac{i \pi}{2} m_a R \, H_1^+(m_a R) J_0(m_a r) & (r < R) \\
\displaystyle - \frac{i \pi}{2} m_a R \, J_1(m_a R) H_0^+(m_a r) & (r > R)
\end{cases}
\,, \label{eq:Esolenoid}\\
B^{\phi}_\mathrm{ind} &= \frac{\pi}{2} \gag a_0 B_0 e^{-im_a t}  m_a R
\begin{cases}
\displaystyle H_1^+(m_a R) J_1(m_a r) &(r < R) \\
\displaystyle J_1(m_a R) H_1^+ (m_a r) & (r > R)
\end{cases} \,. \label{eq:Bsolenoid}
\end{align}
where $J_n(x)$ [$H^+_n(x)$] is the Bessel [Hankel] function of the first kind of order $n$. These results agree with the ones obtained in
Ref.~\cite{Ouellet:2018nfr} by solving the macroscopic equations of
motion for the induced EM field for the same configuration. 

The relevant set of Maxwell's equations are given e.g., in eqs.~(4b) to (4e) of \cite{Berger:2023muj}. Neglecting axion gradients, and using that there are no free charges in the considered configuration they are
\begin{align}
    &\vec\nabla\cdot\vec E =  0 \,,\qquad \vec\nabla\cdot\vec B = 0 \label{eq:maxw1}\\
    &\vec\nabla\times \vec E - \frac{\partial \vec B}{\partial t} = 0 \label{eq:maxw2}\\ 
    &\vec\nabla\times \vec B - \frac{\partial \vec E}{\partial t} =
    \vec J + \gag  \frac{\partial a}{\partial t} \vec B \label{eq:maxw3}
\end{align}
At zeroth order in the coupling $\gag$ the current $\vec J$ in the solenoid generates the constant magnetic field $B_0$. 

Now we need to show that the induced fields satisfy the equations at linear order in $\gag$, with $\vec J = 0$. To this order, in the last term in \cref{eq:maxw3} the magnetic field is the zeroth-order external magnetic field $B_0$ along the $z$-direction and the axion field is given in \cref{eq:axion}. With the divergence in cylindrical coordinates,
\begin{align}
    \vec\nabla\vec V = \frac{1}{r} \frac{\partial}{\partial r}(r V^r)
    +\frac{1}{r} \frac{\partial V^\phi}{\partial \phi} 
+\frac{\partial V^z}{\partial z} \,,
\end{align}
it is apparent that the divergence of both, $\vec E_{\rm ind}$ and $\vec B_{\rm ind}$, vanish and therefore satisfy \cref{eq:maxw1}.
Using the curl in cylindrical coordinates we find the only non-zero components as
\begin{align}
    (\vec\nabla \times \vec E_{\rm ind})^\phi & = - \frac{\partial E^z_{\rm ind}}{\partial r} \,,\\
    (\vec\nabla \times \vec B_{\rm ind})^z & = 
    \frac{1}{r} \frac{\partial}{\partial r}(r B^\phi_{\rm ind}) =
    \frac{B^\phi_{\rm ind}}{r} + \frac{\partial B^\phi_{\rm ind}}{\partial r} \,.
\end{align}
Then it is straightforward to check, that \cref{eq:maxw2,eq:maxw3} are satisfied by using the relevant relations for the Bessel and Hankel functions. 
Focusing on Amp\'ere's law~\eqref{eq:maxw3}, using the induced fields~\eqref{eq:Esolenoid} and~\eqref{eq:Bsolenoid}, we obtain 
\begin{align}
(\vec\nabla\times \vec B_{\rm ind})^z - \frac{\partial \vec E^z_{\rm ind}}{\partial t} = 
\begin{cases}
-i\gag m_a a_0 B_0 \,e^{-i m_a t} & (r < R) \\
0 & (r > R)
\end{cases} \,.
\end{align}
And similarly for \cref{eq:maxw2}:
\begin{align}
    &(\vec\nabla\times \vec E_{\rm ind})^\phi = - \frac{\partial B_{\rm ind}^\phi}{\partial t} \,,
\end{align}
which holds both for $r<R$ and $r>R$.

Hence, we confirm that the induced fields satisfy Maxwell's equations, including Amp\'ere's law.

In \cite{Beutter:2018xfx} we provide also expressions for the ``small experiment'' by expanding the full expressions, \cref{eq:Esolenoid,eq:Bsolenoid}, in the small quantity $(m_aR)$, see eqs.~(3.6) and (3.7) of \cite{Beutter:2018xfx}. These expressions satisfy Maxwell's equations order by order. Note that in eq.~(3.7) of \cite{Beutter:2018xfx} for $B^\phi_{\rm ind}$ there is a sign error~\footnote{We thank J.~Berger and A.~Bhoonah for triggering us to re-check eqs.~(3.6) and (3.7) which allowed us to discover this sign mistake.}. As $E^z_{\rm ind}$ is of higher order in $(m_aR)$ than 
$B^\phi_{\rm ind}$, the latter needs to be expanded up to order $m_a^3$ in order to see explicitly that Maxwell's equations are satisfied at that order.

\section{Conclusions}

In this note we have addressed the critique of BB~\cite{Berger:2023muj} on our previous paper \cite{Beutter:2018xfx}. We have shown that their argument related to the type of used propagator (Feynman versus retarded) is irrelevant for situations of interest, and we do not agree with their second point, stating that our solutions would not fulfill Amp\'ere's law. 

We partially agree with BB that using the Feynman propagator may lead to acausal results in the case of time-dependent classical sources, on time scales of order $t\sim 1/m_a$ or $t\sim R$ (whatever is larger). In this case, more care has to be taken in using the appropriate Greens function and/or combining solutions in order to satisfy temporal boundary conditions. However, in \cite{Beutter:2018xfx} we were interested in stationary solutions, sufficiently far away from effects related to switching on/off the experiment. In these cases, our method provides the correct result which is independent of whether we use Feynman or retarded propagators.

In all cases, we find that the induced electric field is proportional to
\begin{align}
  \vec E \propto m_a^2 R^3 \int_0^Q dq \frac{q^2}{m_a^2 - q^2} \sim \left\{
  \begin{array}{l@{\quad}l}
    m_a^2 R^3 Q = (m_aR)^2 & (Q\gg m_a) \\
    m_a^2 R^3 Q^3/m_a^2 = 1 & (Q\ll m_a)
  \end{array} \right.
\end{align}
where $R=1/Q$ is the size of the experiment. Hence, the result that the induced electric field is parametrically suppressed in a small experiment is confirmed. This statement holds if the measurement time is long compared to $1/m_a$, which is typically the case. In this regime acausal effects due to the use of the Feynman propagator are negligible.

Finally, in \cref{sec:maxwell} we have explicitly shown, that the solutions for the axion-induced electric and magnetic fields obtained in the case of an infinitely long solenoid as obtained in \cite{Beutter:2018xfx} do satisfy Maxwell's equations. We do not expect that the effects of the finite size of wires in a real solenoid would change this conclusion qualitatively. Any corrections due to finite wires of dimension $d$ are expected to be suppressed by some powers of $d/R$. In particular, sufficiently far away from the wires, our idealized solutions should hold with good accuracy. 

\subsection*{Acknowledgement}

We are grateful to Andreas Pargner for useful comments on the manuscript and we thank Joshua Berger and Amit Bhoonnah for communication on this topic.

\bibliographystyle{JHEP_improved}
\bibliography{./refs}

\end{document}